\documentclass[twocolumn,showpacs,preprintnumbers,amsmath,amssymb]{revtex4}

\usepackage{graphicx}
\usepackage{dcolumn}
\usepackage{bm}

\newcommand {\be}{\begin{equation}}
\newcommand {\ee}{\end{equation}}
\newcommand {\bea}{\begin{eqnarray}}
\newcommand {\eea}{\end{eqnarray}}
\newcommand {\bem}{\begin{displaymath}}
\newcommand {\eem}{\end{displaymath}}
\newcommand {\f}{\frac}
\begin{document}

\preprint{ }

\title{Normal zone in $YBa_2Cu_3O_{6+x}$-coated conductors}
\author{ George A. Levin and Paul N. Barnes}
\affiliation{Air Force Research Laboratory, Propulsion Directorate, Wright-Patterson Air Force Base, OH 45433}
%

\date{\today}

\begin{abstract}
We consider the distribution of an electric field in YBCO-coated conductors for a situation in which the DC transport current is forced into the copper stabilizer due to a weak link -- a section of the superconducting film with a critical current less than the transport current. The electric field in the metal substrate is also discussed. The results are compared with recent experiments on normal zone propagation in coated conductors for which the substrate and stabilizer are insulated from each other. The potential difference between the substrate and stabilizer, and the electric field in the substrate outside the normal zone can be accounted for by a large screening length in the substrate, comparable to the length of the sample. During a quench, the electric field inside the interface between YBCO and stabilizer, as well as in the buffer layer, can be several orders of magnitude greater than the longitudinal macroscopic electric field inside the normal zone. We speculate on the possibility of using possible microscopic electric discharges caused by this large ($\sim $kV/cm) electric field as a means to detect a quench.
\end{abstract}
\maketitle

\section{\label{sec:level1}Introduction\protect}
Coated conductors are the subject of intensive studies in recent years due to the potential for commercialization of high-$T_c$ superconductors. The architecture of coated conductors can be described as a "sandwich"  in which the $YBa_2Cu_3O_{7-x}$  (YBCO) superconducting film of about $1 \mu m$ thick is enclosed between a buffered metal substrate of relatively high resistivity ($Ni-5\%W$ alloy, Hastelloy, or stainless steel) and a copper stabilizer layer\cite{Larbalestier,Schoop,Xie,Iijima, Usoskin1}.

Regardless of the method of stabilizer application there is a very thin interfacial layer between the superconducting YBCO and copper that accounts for most or a large fraction of the resistance to the current exchange. This interface resistance is thought to be determined by a few "dead" (underdoped) unit cells of YBCO which has high normal state resistivity in the $c-$direction\cite{Ekin,Angurel}. The current exchange between the superconducting film and normal metal depends on the relationship between the interface resistance and the thickness and resistivity of the stabilizer. If the interface resistance is greater than the transverse resistance of the stabilizer, an exact three dimensional (3D) formulation of the current exchange problem can be reduced to 2D planar model\cite{Levin}.

In this paper we apply the planar approximation in order to better understand some of the recent experimental findings. Specifically, unusual effects were reported in the literature regarding normal zone propagation in coated conductors\cite{Duckworth,Wang}. During a quench a potential difference has been detected between the stabilizer and substrate. A more detailed investigation also revealed that the electric field exists in the substrate well outside the normal zone. A discretized numerical model\cite{Breschi} in which a continuous conductor is modeled as a series of resistors and inductors has been able to reproduce some of the effects. Here we consider these phenomena on the basis of a continuum model which is more physically transparent and is simple enough to allow an analytical solution. The main difference between the electric field distribution in the stabilizer and the substrate is the value of the screening length (also known in literature as the current exchange length). In the stabilizer this length is small in comparison with the length of a typical sample.  In the substrate the screening length is comparable to the length of the samples used in the experiments\cite{Duckworth,Wang}.  

Our results also indicate that there is large electric field, in the range of $kV/cm$, across the interface layer and across the buffer layer. We speculate that a large field across the insulating buffer, which accompanies the quench, might be used as a means to detect the quench by detecting the electromagnetic emissions in the radio-frequency range due to microscopic discharges across the buffer.

This paper is structured as follows: In Section 2 a planar model that describes the voltage distribution in the stabilizer and substrate is derived based on the condition of current conservation. In Section 3 a "weak link" scenario is discussed. This is a situation when there is a section of coated conductor in which the value of the critical current is smaller than the transport current. As a result, a substantial fraction of the transport current is diverted into the stabilizer. We obtain the potential distribution in the stabilizer and the superconducting film within the framework of the Bean model. In Section 4 the potential and electric field in the substrate is discussed in conjunction with the experimental results reported in\cite{Duckworth,Wang}. In Section 5 we consider the implications of the large transverse electric field across the insulating barriers –- the substrate buffer and YBCO-stabilizer interface.

\section{\label{sec:level1}Current exchange within coated conductor\protect}

Let us consider a layer of a normal metal in contact with a superconducting film, Fig. 1. 
Let $\bar {R}$ (with dimensionality $\Omega\; cm^2$) be the  resistivity of the interface between them. Here we consider only DC electric field and currents.
The 2D density of current
\be
\vec{J}_{1,2}=-\f{d_{1,2}}{\rho_{1,2}}\nabla V_{1,2},
\ee
where $V(x,y)$ is the electric potential, $d_i$ is the thickness, and $\rho_i$ is the resistivity of a given metal layer.  The subscripts $1$ and $2$ refer to the stabilizer and the substrate respectively.  For the stabilizer the current conservation takes form:
\be
\oint (\vec{J}_1\cdot \vec{n})d\ell -\int j_z dA=0
\ee
Here the integration is carried out over the length and area of an arbitrary 2D contour; $\vec{n} $ is the outward directed unit vector, and $j_z$ is the density of current flowing across the interface between the superconducting film and the normal metal. Taking into account Eq. (1) we obtain
\bea 
\nabla\cdot\vec{J}_1-j_z=0;\; \\ \nonumber
\f{d_{1}}{\rho_{1}}\Delta V_1+j_z=0
\eea
If we assume an ohmic relationship between the current density $j_z$ and the potential difference between the stabilizer and the superconductor
\be
j_z=-\frac{V_1-V_s}{\bar{R}_1},
\ee
we get the 2D Helmholtz equation
\be
\Delta V_1 -\kappa_1^2 (V_1-V_s)=0.
\ee
Here $\kappa_1 \equiv \lambda^{-1}_1 $ defines a screening length 
\be
\lambda_1 =\left (\frac{ d_1\bar {R}_1 }{\rho_1 }\right )^{1/2}.
\ee
\begin{figure}
\includegraphics{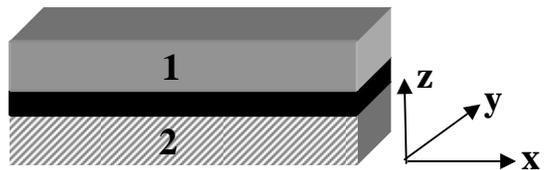}
\\
\caption{\label{fig:} A sketch of a coated conductor (not to scale). A thin superconducting film is sandwiched between the copper stabilizer (1) and metal substrate (2). The resistive interface between copper and $YBCO$, as well as the insulating buffer between $YBCO$ and substrate are not shown, because they are much thinner than even the $YBCO$ film.}
\end{figure}

The 2D approximation defined by Eqs.(1) -- (3) is valid as long as $d_1\ll a_1$, 
where $a_1$ is a characteristic length scale\cite{Levin}:
\be
d_1\ll a_1 \equiv \frac{\bar{R}_1}{\rho_1}.
\ee

Similarly, we can obtain the 2D Helmholtz equation that describes the potential in the substrate:
\be
\Delta V_2 -\kappa^2_2 (V_2-V_s)=0.
\ee
The screening length $\lambda_2 \equiv \kappa^{-1}_2 $ is determined by the resistance of the insulating buffer between YBCO and substrate. 
\be
\lambda_2 =\left (\frac{ d_2\bar {R}_2 }{\rho_2 }\right )^{1/2}.
\ee

In the stabilizer the screening length is much smaller ($\lambda_1\sim 300\mu m$, $\bar {R}_1\sim 50 n\Omega cm^2$) than the other dimensions of the problem (length and width of the conductor)\cite{Polak,Levin}. In contrast, the superconductor film in the coated conductor is grown on an insulating buffer about $150\;nm$ thick. The buffer serves as a highly resistive interface between the substrate and YBCO. At liquid nitrogen temperature the resistivity of the substrate, either $Ni$-$5\%W$ alloy, Hastelloy, or stainless steel is two to three orders of magnitude higher than that of copper, but the interface resistivity $\bar{R}_{2}$ determined by the buffer is also many orders of magnitude greater than that between copper and YBCO. Thus, the screening length $\lambda_2$ (Eq. (9)) in the substrate is several orders of magnitude greater than that in copper. 

The potential $V_s$ in the superconducting film is determined by the equation of conservation of the two-dimensional density of current $\vec{J_s}\;(A/cm)$\cite{Levin}
\be
\nabla\cdot\vec{J_s}=- j_{1}+ j_2 =\frac{{V_1}-V_s}{\bar{R}_1} + \frac{{V_2}-V_s}{\bar{R}_2}.
\ee
Here $j_{1,2}$ are the densities of current flowing across the interfaces in and out of stabilizer and substrate respectively, see Eqs. (3) and (4). 

A constituent relation for the superconductor can be written as $\vec{E}\equiv r(J_s)\vec{J_s}$, which translates into a non-linear equation for the potential:
\be
\nabla [r^{-1}(|\nabla V_s|)\nabla V_s]=-\frac{{V_1}-V_s}{\bar{R}_1}
-\frac{{V_2}-V_s}{\bar{R}_2}.
\ee
For power law dependence\cite{Friesen}
\be
r(J_s)=\f{E_0}{J_c}\f{|J_s|^{n-1}}{|J_c|^{n-1}},
\ee
Eq. (11) takes form:
\be
\nabla \left [\nabla V_s \left (\f{|\nabla V_s|}{E_0}\right )^{(1-n)/n}\right ]=\Lambda_1^{-2}(V_1-V_s)+\Lambda_2^{-2}(V_2-V_s).
\ee
Here $J_c$ is the critical current density and $E_0=1 \mu V/cm$. The two screening lengths in the superconductor 
\be
\Lambda_{1,2} =\left (\f{\bar{R}_{1,2}J_c}{E_0}\right )^{1/2},
\ee
along with those in the stabilizer and substrate - Eqs.(6) and (9), determines the distances over which the current exchange between the superconductor, stabilizer, and substrate takes place. As long as $\Lambda_2\gg\Lambda_1$, the current exchange between YBCO and substrate does not influence the potential distribution in either the superconducting film or the stabilizer. Correspondingly, the respective terms in the right-hand side of Eqs. (11) and (13) can be omitted. On the other hand, the critical current density in coated conductors $J_c\sim 200\; -\; 400 A/cm$, so that even the smallest of the two screening lengths in the superconductor
\be
\Lambda_1\sim 2\;-\;3 \;cm.
\ee
This is much greater than that in the stabilizer
\be
\lambda_1\sim 0.3\; mm. 
\ee
Thus, the effects of the current exchange in the stabilizer manifest itself over much shorter distances than those in the superconductor. The potential difference between superconductor and stabilizer tends to be eliminated over the distance of the order of $\lambda_1$ because the stabilizer adjusts its potential to that of the superconductor. 

The length scales of the potential distribution in the superconducting films tend to be large, of the order of $bn$ or $bn^{1/2}$, where $b$ is the size of a defect or an obstacle to current and $n\sim 20-40$ is the exponent in Eq. (12)\cite{Friesen}. Therefore, the details of the current exchange that takes place on the spatial scale of the order of $\lambda_1$ in many situations can be accurately described by assuming that the potential of the superconductor is constant or piecewise constant\cite{Levin}. 

\section{\label{sec:level1} Normal zone: Weak link scenario}

In this section we consider the situation when stabilization has failed and the normal zone starts to spread along the conductor. We will not consider the temporal development of the normal zone, but limit ourselves to a question of instantaneous distribution of electric field in coated conductor, given a certain position and length of the normal zone. The normal zone boundaries can be defined simply as points at which the transport current $I>I_c(T)$, where $T$ is the local temperature of the superconductor. The ``supercritical`` regime - the regime when the superconductor carries current approximately equal $I_c$ and the substantial current $I-I_c$ flows through the stabilizer can be described by Eq. (5). However, the superconducting film cannot be treated as equipotential. 

As an example, let us consider a simple model of a ``weak link``, where the transport current $I$ flows in a coated conductor with non-uniform critical current. Namely, outside the weak section of length $l$, at $|x|>l/2$, the critical current is greater than $I$. Here the origin is chosen in the center of the weak section. The exact value of critical current outside the normal zone associated with the weak section it is not important. In the central section $|x|<l/2$ the critical current $I_c<I$. As the result, part of the total current will be forced into the stabilizer. 
  
The electric field in the superconductor increases very rapidly when the transport current exceeds the critical value:
\be 
E=E_0(I/I_c)^n,
\ee
where $E_0$ is defined as $1\mu V/cm$ and the exponent $n$ is large ($n\approx 20-\;40$). The Bean  model is the limit $n\rightarrow\infty$. In this limit, which we adopt hereafter in order to obtain an analytical solution, the superconductor can generate arbitrarily large electric field. Therefore, the potential of the superconductor is not required to be continuous across the discontinuity at the critical current. The potential and electric field in the stabilizer must be continuous.

The potential of the superconducting film outside the normal zone, where the critical current is greater than the transport current, we take to be constant:
\be
V_s=\pm \delta V/2,
\ee
where $\delta V$ is the total potential drop across the conductor, and the signs $\pm$ refer to $x>l/2$ and $x<-l/2$ respectively. This approximation means that we neglect the right-hand side of Eq. (13) due to the large values of $\Lambda_i$\cite{Levin}. 
\begin{figure}
\includegraphics{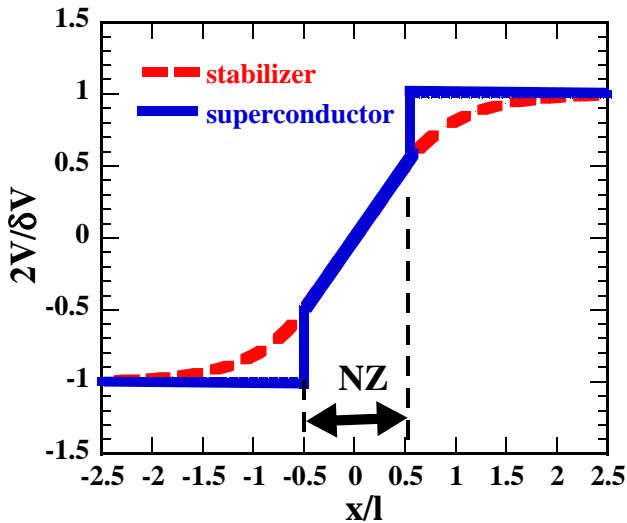}
\\
\caption{\label{fig:} (Color online)Potential in the superconducting film and stabilizer, Eqs(18) and (21)-(23). For this plot we chose the value of $\lambda_1=l/2$. The location and width of the normal zone (NZ) is indicated. The voltage is normalized to $\delta V/2$ and coordinate is normalized to the length of the normal zone $l$. }
\end{figure}

A solution of Eq. (5) we chose in the form:
\be
V_1=\pm \delta V/2 + Ae^{-\kappa_1 |x|};\;  |x|>l/2.
\ee 
Inside the normal zone, at $|x|<l/2$, the current in the superconductor cannot exceed $I_c$ and, therefore, the current $I-I_c$ must flow through the stabilizer generating electric field 
\be
E_c=\f{\rho_1 (I-I_c)}{Wd_1}. 
\ee
Correspondingly, the potential of the stabilizer for $|x|<l/2$ has the form 
\be
V_1=E_cx.
\ee
The solution (21) can satisfy Eq. (5) only if the potential of the superconductor is the same as the stabilizer (no current exchange between superconductor and stabilizer):
\be
V_s=E_cx;\;\; |x|<l/2.
\ee
This condition can be understood as follows:  In the Bean model an infinitesimally small variation of current around the value of $I_c$ can cause finite change in electric field. Therefore, any potential difference between the superconductor operating in the critical regime and an adjacent normal metal can be eliminated by a negligible amount of current transferred between them.
The requirement of continuity of $V_1$ and $\partial V_1/\partial x$ at $x=\pm l/2$ determines the unknown parameters $A$ and $\delta V$. Thus,
\be
V_1=\pm \left ( \f{\delta V}{2}- E_c\lambda_1 e^{-\kappa_1(|x|-l/2)}\right )          ;\;\; |x|>l/2,
\ee
where $\delta V $ is the potential difference required to maintain current $I>I_c$ across the weak link:
\be
\delta V=E_c(l+2\lambda_1)=\f{\rho_1 (l+2\lambda_1 )}{Wd_1}(I-I_c).
\ee
Figure 2 illustrates the potential distribution in the superconducting film and the stabilizer. As shown, the length of the normal zone is chosen to be twice the screening length: $l=2\lambda_1$. 
Notice that the potential and electric field in the superconductor are discontinuous at the boundaries of the normal zone. The potential of the superconducting film jumps by the amount $E_c\lambda_1$ at $x=\pm l/2$. 
This is the consequence of the Bean model approximation. Current exchange between stabilizer and superconducting film takes place outside the normal zone, where there is potential difference across the interface. 
The total power dissipated by the conductor in the stabilizer, superconducting film and interface
\be
Q=\delta V I =\f{\rho (l+2\lambda )}{Wd}I(I-I_c).
\ee

\section{\label{sec:level1} Electric field in the substrate \protect}

Experiments dealing with normal zone propagation in coated conductors have uncovered some rather peculiar features; the electric field in the copper stabilizer is essencially localized within the normal zone, because the screening length in currently manufactured coated conductors is $\lambda_1\sim 0.3 mm$ and, therefore, $\lambda_1\ll l$. In the substrate, however, the electric field is detectable over the total length ($12\;cm$) of the sample\cite{Wang}. Also, in similar samples the stabilizer and substrate are found to not be equipotential\cite{Duckworth}. 

These phenomena can be readily understood within our schematic model of the coated conductor as shown in Fig. 1. The stabilizer and the substrate have no direct contact.  The potential in the substrate $V_{2}$ is subject to Eq. (8) with the screening length given by Eq. (9).
To our knowledge, the interface resistivity between the substrate and superconductor has never been measured, but from experiments discussed below it is obvious that in practical coated conductors the screening length in the substrate $\lambda_2$ either greatly exceeds or, at least, on the order of 10 cm.  In order to understand the results of the experiments where the length of the samples $L$ is of the order of the screening length $\lambda_2$, we consider the solutions of Eq. (8) in the limit $\kappa_2L\sim 1$ or $\kappa_2L\ll 1$. 

The boundary conditions for the potential in the substrate are determined by the requirement that the electric field (and current density) vanishes at the ends of the sample. 
\be
E_x\big |_{x=\pm L/2}=0.
\ee 
Here $L$ is the length of the sample. For simplicity, we will continue to consider a symmetrically spaced normal zone of length $l$ located at $-l/2<x<l/2$. The potential of the superconducting film at the edges $x=\pm L/2$ is $V_s=\pm\delta V/2$. Because of this symmetry it is sufficient to discuss the solution only at $x>0$.

A solution of Eq. (8), subject to the boundary condition (26) is given by 
\be
V_{2}-V_s=A\cosh \{\kappa_2(x- L/2)\};\;\; l/2<x<L/2;\;\;\ \\  
\ee
Because of the high resistivity of the buffer, the substrate inside the normal zone ($|x|\leq l/2$) is not necessarily equipotential with the superconducting film. The stabilizer and the superconducting film are still equipotential in this area and, therefore, the potential in the superconducting film is given by Eq.(22): $V_s=E_cx$. Thus, for $-l/2\leq x\leq l/2$, the Eq. (8) for the substrate potential has the form:
\be
\Delta V_{2} -\kappa^2_2 V_{2}=-\kappa^2_2 E_cx,
\ee
with a  solution
\be
V_{2}=B\sinh\{\kappa_2x\}+ E_cx.
\ee
The coefficients $A$ and $B$ are determined by matching the solutions (27) and (29) at $x=l/2$:
\bea
\delta V/2 +A\cosh \{\kappa_2((l- L)/2)\}= 
B\sinh\{\kappa_2l/2\}+\;\; \\\nonumber +E_cl/2;\\\nonumber
\kappa_2 A\sinh \{\kappa_2((l- L)/2)\}= \kappa_2 B\cosh \{\kappa_2l/2\}+ E_c.
\eea

The relationship between the electric field in the stabilizer $E_c$ and the potential difference $\delta V$ is given by Eq. (24). For simplicity we will consider a situation in which the screening length in the copper stabilizer is much smaller than the length of the normal zone, so that $\delta V\approx E_cl$. Introducing the dimensionless coordinate and parameters:
\bea
x^{\prime}\equiv 2x/L;\;\;
\gamma \equiv \kappa_2L/2;\;\;\xi \equiv l/L, 
\eea
we obtain from the first of Eqs. (30):
\be
A\approx B\f{\sinh\{\gamma\xi \}}{\cosh\{\gamma (1-\xi)\}}.
\ee

As the result, the Eq. (27) takes form:
\bea
V_{2}=\f{\delta V}{2} \left (1-\f{1}{\gamma \xi } \f {\sinh\{\gamma\xi\} }
{\cosh\{\gamma \}}\cosh\{\gamma (1-x^{\prime})\}\right );\\
\;\; \xi <x^{\prime}<1. \nonumber
\eea
\begin{figure}
\includegraphics{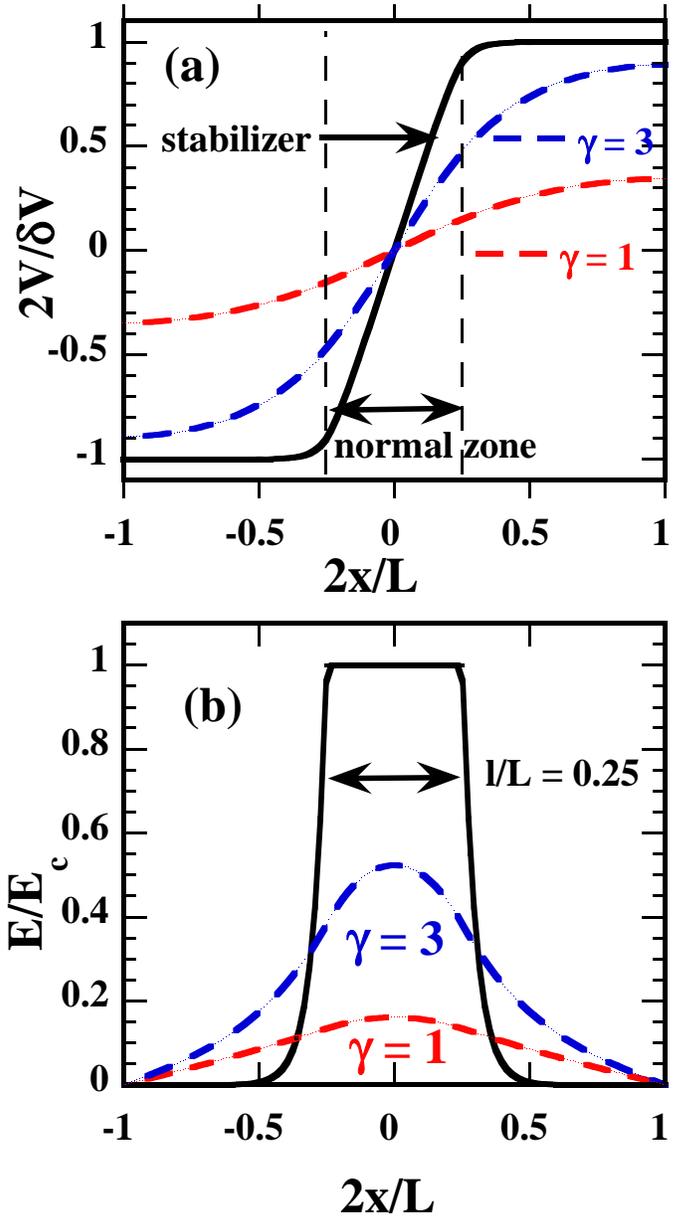}
\caption{\label{fig:} (Color online) (a) Electric potential in the substrate (dashed curves) and in the stabilizer (solid curve) for $\gamma =1$ and $\gamma = 3$. The length of the normal zone $l= 0.25 L$.  (b) The electric field in the substrate (dashed curves) and in the stabilizer (solid curve) for the same parameters as in Fig. 3(a). In both figures the double arrow shows the location and length of the normal zone.}
\end{figure}

The solution (29) takes form:
\bea
V_{2}=\f{\delta V}{2} \left (\f{x^{\prime}}{\xi }-\f{1}{\gamma \xi } 
\f {\cosh\{\gamma (1- \xi)\} }
{\cosh\{\gamma \}}\sinh\{\gamma x^{\prime}\}\right );\\
|x^{\prime }|\leq \xi. \nonumber
\eea

The electric field in the substrate $|E_{2}|= |\partial V_{2}/\partial x|$ is given by 
\bea
|E_{2}|= E_c \f {\sinh\{\gamma\xi\} }
{\cosh\{\gamma \}}\sinh\{\gamma (1-x^{\prime})\};\;\; \xi <x^{\prime}<1.\;\;\; \\
|E_{2}|= E_c \left [1- 
\f {\cosh\{\gamma (1- \xi)\}}
{\cosh\{\gamma \}}\cosh\{\gamma x^{\prime}\}\right ];
|x^{\prime }|\leq \xi. \nonumber
\eea
Here $E_c\approx \delta V/l$ is the electric field in the stabilizer inside the normal zone, Eq.(20). 

Figure 3(a) shows the distribution of the electric potential in the substrate (Eqs. (33), (34)) for two values of $\gamma $: $\gamma =3$ and $\gamma =1$, and $\xi =0.25$ (Eq. (31)). For comparison the potential of the stabilizer, similar to that shown in Fig. 2, is also included. The potentials in the figure are normalized to $\delta V/2$ (Eq. (24). 
The length of the normal zone is taken to be $1/4$ of the length of the sample. Even for relatively short screening length ($\gamma =3$ corresponds to $\lambda_2 =L/6 $) there is a noticeable potential difference between stabilizer and substrate. This potential difference reaches maximum near the boundaries of the normal zone. 
 
In Fig. 3(b) the electric field in the substrate is shown (Eqs. (35)). The magnitude of the field is normalized with $E_c$ (Eq. (20)), which is the field inside the stabilizer in the area of the normal zone. The electric field in the stabilizer is localized within the area of the normal zone, but in the substrate it is spread out well outside the normal zone. 

In Fig. 4 the variation of the field with the length of the normal zone is illustrated. The magnitude of the field in the substrate monotonically increases with the length of the normal zone. Thus, if there are two pairs of electrodes attached to the stabilizer and the substrate, respectively, in similar positions \cite{Wang,Duckworth}, the potential difference between the voltage taps attached to the stabilizer $V_{stab}$ will stop increasing when the normal zone expands past these electrodes. On the substrate side, the potential drop $V_{sub}$ will continue to increase even when these electrodes are entirely within the normal zone. Of course, the increasing temperature of the stabilizer may result in some increase of the voltage drop on the stabilizer side due to increasing resistivity, but the ratio $V_{sub}/V_{stab}$ will continue to increase, reflecting the spread of the normal zone. 

\begin{figure}
\includegraphics{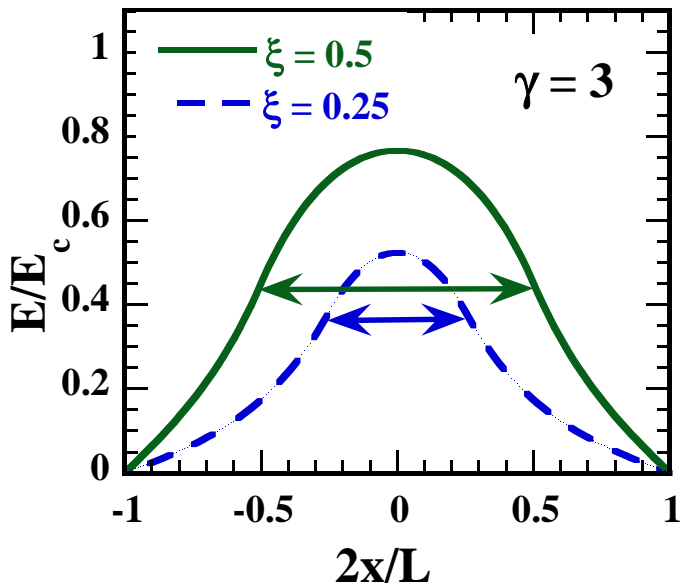}
\caption{\label{fig:} (Color online) The electric field in the substrate for different lengths of the normal zone. The dashed curve corresponds to $l=0.25 L$, same as in Fig 3. The solid curve corresponds to $l=0.5 L$. The double arrows show the locations and lengths of the normal zone. The value of $\gamma =3$ is the same for both curves.}
\end{figure}

As Figs. 3 and 4 demonstrate, when $\lambda_2 \sim L$ the distribution of the electric field and potential in the substrate exhibit no clear indication where the boundaries of the normal zone are. This is precisely what Wang {\it et al}.\cite{Wang} have reported after initiating quench in a $12\;cm$ long sample. If, however, the screening length $\lambda_2$ greatly exceeds the sample length ($\gamma \ll 1$), the electric field inside the substrate becomes very small (see Eq. (35) and a trend obvious in Fig. 3(b)). Therefore, the experimental results reported in Ref.\cite{Wang} indicate that  $\lambda_2\sim L\sim 10\;cm$. From this observation we can estimate the buffer resistance 
\bea
\bar{R}_{2}=\f{\lambda_2^2\rho_{2}}{d_{2}}\sim 0.4 \;\Omega\;cm^2.\nonumber
\eea
Here we took $d_{2}=75\;\mu m$ and $\rho_{2}\sim 20 -  30 \;\mu\Omega\;cm$ ($Ni-5\%W$ at $77\;K$)\cite{Wang}. The resistivity of the buffer material
\bea
\rho_{b}\sim \f{\bar{R}_{2}}{d_{2}}\sim 2\times 10^4\;\Omega\;cm ,\nonumber
\eea
where the thickness of the buffer $d_{b}$ is taken to be $200\; nm$
\cite{ Schoop}. 

The electric field in the substrate is smaller, but of the same order of magnitude as that in the stabilizer, Fig. 3(b). However, the resistance of the substrate is two to three orders of magnitude higher than that of the stabilizer. Correspondingly, the total current flowing through the substrate is also much smaller than that in the stabilizer:
\be
I_{1}=\f{wdE_c}{\rho_1};\;\; I_{2}=\f{wd_{2}E_{2}}{\rho_{2}}; \;\;I_{1}\gg I_{2}.
\ee
Here $I_{1}$ and $I_{2}$ are the currents in the stabilizer and substrate respectively ($I_{1}+I_{2}=I-I_c$). Considering that in coated conductors $d_1\sim d_{2}$ and $\rho_1 \ll \rho_{2}$, it is obvious that the substrate carries a negligible fraction of the total current. However, the electric field outside the normal zone is detectable only in the substrate and, as long as $\lambda_2\sim L$, its magnitude is of the order of that in the stabilizer inside the normal zone. 

\section{\label{sec:level1} Electric field across the buffer and interface\protect}

As Fig. 3(a) illustrates, there is a substantial potential difference between the substrate and the superconducting film that reaches maximum at the boundary of the normal zone. Since the substrate is metallic, this potential drop falls entirely across the insulating buffer segregating the YBCO film and the substrate. The buffer, $150 - 200\; nm$ thick, consists of several layers of insulating compounds ($Y_2O_3, \;CeO_2$, etc.)\cite{Schoop,Xie}. Therefore, the electric field across the buffer $E_{\perp}$ may be large even for relatively small values of the potential difference across the normal zone $\delta V$. For example, the electric field in the stabilizer inside the normal zone, Eq.(20):
\be
E_c\sim \rho J/d\sim 10^{-2}\; V/cm.
\ee
Here we took a linear density of current in the stabilizer $J=(I-I_c)/W\sim 100\;A/cm$ and the sheet resistance of copper stabilizer $\rho_1 /d_1\sim 10^{-4}\;\Omega$. Correspondingly, for the normal zone of length $l$, the electric field across the buffer of thickness $d_b$:
\be
|E_{\perp}|\sim\f{\delta V}{d_{b}}\sim E_c\f{l}{ d_{b}}\gg E_c.
\ee
For a $2\;cm$ long normal zone and $d_b\sim 200\;nm$, $ E_{\perp}\sim 10^3\;V/cm$.
The maximum of the potential difference between the substrate and the superconducting film is reached at the edges of the normal zone, Fig. 3(a). Thus, the spread of the normal zone is accompanied by the emergence of an electric field across the insulating buffer with a magnitude in the kV/cm range. The dielectric breakdown strength of the buffer is not known, and it is possible that under certain conditions microscopic electric discharges may take place in the buffer during quench. 

A similar phenomenon $-–$ existence of large transverse electric field across the interface between the stabilizer and superconducting film $–-$ follows from our analysis of the potential in the stabilizer, see Eq. (23) and Fig. 2. Here, the maximum potential difference across the interface $|V_1-V_s|=E_c\lambda_1$ at $x=\pm l/2$. In the planar model, Eqs. (3)$-–$(5), we treat the interface as an infinitesimally thin layer that can have a finite potential drop across it. Obviously, in reality it has a finite, submicron thickness $d_{int}$ and a certain average resistivity $\rho_{int}$. The transverse electric field inside the interface
\be
|E_{int}|\sim E_c\f{\lambda_1}{ d_{int}}\gg E_c.
\ee
Equation  (4) can be written as
\bea
\left.  j_z\right |_{z=0}=\f{E_{int}}{\rho_{int}}\equiv 
\f{E_{int}d_{int}}{\rho_{int} d_{int}}\equiv -\frac{{V}-V_s}{\bar{R}_1}.
\eea
A very rough estimate of $d_{int}$ can be obtained assuming that the resistivity of the interfacial material is of the order of the out-of-plane normal state resistivity $\rho_c$ of strongly underdoped (and, therefore, nonsuperconducting at liquid nitrogen temperatures) $YB_2Cu_3O_{6+y}$. For $y< 0.5$ the resistivity in the out-of-plane direction $\rho_c\sim 0.1 \;\Omega \;cm$ at $T\approx 77\;K$\cite{Cimpoiasu,Zverev}. Therefore, according to Eq. (40),
\bea
d_{int}=\f{\bar{R}_1}{\rho_c}\sim \f{50\; n\Omega\;cm^2}{0.1\; \Omega\;cm}\sim 5\;nm .\nonumber
\eea
In YBCO the size of the unit cell in the $c-$direction is approximately $ 1.2\;nm$. Thus, just a few unit cells of underdoped YBCO can account for the value of interface resistivity in coated conductors.

The electric field across the interface can be very large, and it increases in proportion to $\lambda_1$. For a relatively high interface resistivity $\bar{R}$ it can be on the order of  $1 \;kV/cm$ or even greater. It is therefore likely that the commonly assumed ohmic relationship between current and voltage across the interface (Eq. (4)) is not accurate and can be used only as a rough approximation. If the interface resistivity is field dependent (decreases with increasing field)  $\bar{R}=\bar{R}(|E_{int}|)$, the exponential variation of the potential in the zone of current exchange, Eqs. (19) and (23) are no longer valid, at least in the area of the largest field $|E_{int}|$. More importantly, the power loss is no longer equally split between the stabilizer and the interface\cite{Levin}. Greater amount of energy will be dissipated in the interface, than in the stabilizer. 

This important point can be illustrated by assuming a constituent relationship different from that given by Eq.(4). Let us take a polynomial dependence of $j_z$ on the potential difference, characteristic of electric discharges in insulators:
\be
j_z=-\frac{(V_1-V_s)| V_1-V_s |^k}{\bar{R}_1U^k},
\ee
where $U>0$ and $k\ge 0$ are arbitrary parameters. The one-dimensional equation (3) takes form
\be
\f{d_{1}}{\rho_{1}}V_1^{\prime\prime} -\frac{(V_1-V_s)| V_1-V_s |^k}{\bar{R}_1U^k}=0.
\ee
In the area of conductor where the superconducting film is equipotential, $V_s=const$, both sides of Eq. (42) can be multiplied by $V_1^{\prime}$, which gives us the following:
\be
\f{\partial }{\partial x}\left (\f{1}{2} \f{d_{1}}{\rho_{1}}(V_1^{\prime})^2 -\f{1}{k+2}\f{(V_1-V_s)^{k+2}}{\bar{R}_1U^k} \right )=0
\ee
The constant of integration is zero because at a distant point the electric field in the stabilizer $V_1^{\prime} \rightarrow 0$ and the stabilizer is equipotential with the superconductor, $V_1-V_s\rightarrow 0$. The areal density of power dissipation\cite{Levin} in the stabilizer is
\be
Q_{st} = \f{d_{1}}{\rho_{1}}(V_1^{\prime})^2.
\ee
The areal density of  power dissipation in the interface
\be
Q_{int}=-j_z(V_1-V_s)=\f{(V_1-V_s)^{2+k}}{\bar{R}_1U^k}.
\ee
Thus, 
\be
Q_{int}=\f{k+2}{2}Q_{st}
\ee
For $k>0$, which is characteristic of discharge phenomena, the power dissipation in the interface, in immediate proximity to the YBCO film, exceeds that in the stabilizer. This may have a significant consequences for the dynamics of quench.

\section{\label{sec:level1} Summary and speculation.}

The anomalies of the electric field distribution in the substrate during quench\cite{Wang,Duckworth} can be accounted for by a large value of the screening length in the substrate, comparable to the length of the sample. As the result, an electric field and relatively small longitudinal current exist in the substrate well outside the normal zone. A significant potential difference between the superconducting film and substrate may result in a large electric field, perhaps as large as few $kV/cm$, across the substrate's buffer, inside and outside of the normal zone. It should be noted that none of the previous superconducting wires –- neither low-$T_c$, nor the first generation high-$T_c$ wires - had such an element as a metal substrate insulated from the superconducting material by high resistivity barrier. These phenomena are specific to coated conductors with on-top stabilizer and their study does not have a long history.

In principle, the strong electric field in the buffer can be easily suppressed, if necessary, by electrical connection between the stabilizer and substrate, as in the surround stabilizer architecture. However, it may be possible to put this phenomenon to use, rather than simply discard it. Quench protection of the coated conductors is an outstanding issue of concern, primarily due to slow propagation of the normal zone. In order to implement active protection schemes, detection of a quench is necessary. An appealing scheme of detection would be one that does not require numerous voltage taps inserted into the coil. For stationary magnets the detection of acoustic signals resulting from quench-induced mechanical disturbances is one of such methods\cite{Iwasa}. Another, perhaps distant, possibility of normal zone detection in real time would be detection of electromagnetic emission in the radio frequency range due to microscopic electric discharges across the buffer. In spite of screening of such emissions by the stabilizer and substrate, the emissions might reach a sensitive antenna inside the magnet from the edges of the tape-like conductor.


.\\

\newpage
\end{document}